\documentclass[12pt]{article}
\usepackage{amsmath}
\usepackage{amssymb}
\usepackage[dvips]{graphicx}

\input epsf

\numberwithin{equation}{section}

\newcommand{\CF}{ {\cal F} }
\newcommand{\CG}{ {\cal G} }
\newcommand{\CN}{ {\cal N} }
\newcommand{\CZ}{ {\cal Z} }

\newcommand{\hv}{ h^{\vee}  }
\newcommand{\av}{ \alpha^{\vee}  }
\newcommand{\AD}{ A_{D}  }

\begin{document}

\addtolength{\baselineskip}{2pt}
\thispagestyle{empty}

\vspace{2.5cm}


\begin{center}
{\scshape\large 
Discrete symmetry in
\\
 supersymmetric $\CN=2$ gauge theory}

\vspace{1cm}

{\scshape\large  Michael Yu. Kuchiev}

\vspace{1.5cm}
{\sl School of Physics, University of New South Wales,\\
Sydney, Australia}\\
{\tt kmy@phys.unsw.edu.au}\\\vspace{1cm}

{\Large ABSTRACT}
\vspace{0.3cm}

\end{center}

A new discrete symmetry group,  which governs low-energy properties of the supersymmetric $\CN=2$ gauge theory is found.  
Each element of this group $S_r$, $r$ being the rank of the gauge group,
represents a permutation of $r$ electric charges available in the theory accompanied by a simultaneous permutation of $r$ monopoles, provided the sets of charges and monopoles are chosen properly. Properties of the theory are strongly influenced by $S_r$; if the central charges (and masses) of $r$ monopoles are degenerate, then the central charges (and masses) of $r$ electric charges are also necessarily degenerate, and vice versa. This condition uniquely defines the vital value of the VEV of the scalar field, at which all monopoles are massless. The general theoretical discussion is illustrated by a model, which generalizes the Seiberg-Witten treatment of the supersymmetric $\CN=2$ gauge theory for an arbitrary gauge group.
\newpage

\date{today}

\section{Introduction}
\label{intro}

The low-energy properties of the supersymmetric $\CN=2$ gauge theory are shown to comply with 
a new discrete symmetry, which finds its origin in the Dirac-Schwinger-Zwanziger quantization condition for electric and magnetic charges of dyons. In simple physical terms it can be described as a group of permutations $S_r$, $r$ being the rank of the gauge group, of  $r$ simple electric charges and $r$ simple magnetic monopoles 
(simple electric charges and monopoles have the electric or magnetic charges, which equal the simple coroots of the Cartan algebra). The existence of this symmetry implies that if central charges of $r$ simple monopoles are equal, then  central charges of $r$ simple electric charges are also equal, and vice versa. Due to geometrical associations this phenomenon is called the Weyl vector alignment.  This property completely defines  parameters that specify the vital state of the theory, in which all monopoles are  massless. The general theoretical treatment is illustrated by the recently proposed model, which is found to comply with  $S_r$.

Seiberg and  Witten \cite{Seiberg:1994rs,Seiberg:1994aj}  suggested that the low-energy description of the $\CN=2$ supersymmetric gauge theory can be presented explicitly and proposed a model, which implemented this idea for the SU(2) gauge group.
Their approach was generalized to cover other gauge groups. In \cite{Klemm:1994qs,Argyres:1994xh} the model for SU($n$) gauge groups was suggested. A brief review and references to other works, in which models for other classical gauge groups were developed are given in \cite{D'Hoker:1999ft,Wu2006503}. Additional references can be found in \cite{Kuchiev:2008mv},
where it was noted that the absence of the universal description presented an important issue as different models and techniques were employed even for theories based on the classical gauge groups; for exceptional groups the situation looked even more difficult. Ref. \cite{Kuchiev:2008mv} also described a special set of ``light" dyons. Each of them can be made massless if the state of the theory is chosen properly. Relying on this fact
Ref. \cite{Kuchiev:2009ez} proposed a general model, which is applicable to the theory based on any compact simple gauge group, classical or exceptional, reproduces known discrete symmetries of the theory and complies with the monodromies at weak and strong coupling.

The present work suggests looking at the low-energy properties of the $\CN=2$ gauge theory
from a more general, model-independent perspective using for this purpose the discrete symmetry available in the theory. The idea is to focus attention on those transformations of the scalar field, which do not change the set of central charges for all dyons available in the theory. As mentioned, this approach reveals that there exists a discrete symmetry $S_r$, which defines vital properties of the theory for $r\ge 2$. For the SU(2) gauge theory the group in question is trivial, $r=1$, $S_1=1$. 


\section{Basic low-energy properties}
\label{Dyons}
Let us outline the known properties of the low-energy approach to
the supersymmetric $\CN=2$ gauge theory that are relevant for our discussion.
The theory is governed by the vacuum expectation value (VEV) of the scalar field, which belongs to the Cartan subalgebra of the gauge algebra. Consequently it can be presented as an $r$-dimensional vector $A$, where the rank $r$ of the gauge group defines the dimension of the Cartan algebra; similarly the VEV of the dual scalar field is also an $r$-dimensional vector $A_D$. This makes the electric $q$ and magnetic $g$ charges of dyons $r$-dimensional vectors as well. 

It is known, see e. g.  \cite{Weinberg:2006rq}, that in order to satisfy the Dirac-Schwinger-Zwanziger quantization condition
the electric and magnetic charges should belong to the lattice defined in the Cartan algebra.
Generally speaking the electric charges should  belong to the lattice 
of roots ${\mathbb Q}$,  while magnetic ones to the lattice of coroots ${\mathbb Q}^\vee$ of the Cartan algebra. However, in the supersymmetric $\CN=2$ gauge theory the discrete chiral transformations, see below, ensure that electric and magnetic charges should belong to the same lattice. Consequently both sets of charges have to occupy vertexes of the same lattice ${\mathbb Q}^\vee$
\begin{equation}
q\,=\,\sum_{i=1}^r\,n^{(q)}_i\,\av_i~,
\quad\quad
g\,=\,\sum_{i=1}^r\,n^{(g)}_i\,\av_i~.
\label{g=ng}
\end{equation}
Here $\av_i$ are the simple coroots of the Cartan algebra from which the lattice ${\mathbb Q}^\vee$ is constructed, and $n^{(q)}_i$ and $n^{(g)}_i$ are integers, 
$n^{(q)}_i,\,n^{(g)}_i\in \mathbb Z$. The two lattices ${\mathbb Q}^\vee$ and $\mathbb Q$ are different only for non-simply laced 
groups when the first one is a sublattice of the second, ${\mathbb Q}^\vee\subset \mathbb Q$,  while for simply laced ones they coincide, ${\mathbb Q}^\vee = \mathbb Q$.
For basic properties of the Lie algebras, which are used in this work see e. g. books  \cite{Bourbaki:2002,Di-Francesco:1997}.
The scalar products $\av_i\cdot \av_j\in \mathbb{Z}$ are integer-valued, 
here and below the dot-product is taken in the Cartan subalgebra. As a result scalar products for any electric and magnetic charges  satisfying (\ref{g=ng}) are also integer-valued, $q\cdot g\in \mathbb Z$, in compliance with the quantization condition.  

The dyons are the BPS states and consequently, as was found by Witten and Olive \cite{Witten:1978mh}, the mass $m_{\CG}$ of a dyon is related to its central charge $\CZ_{\,\CG}$
\begin{align}
 m_{\,\CG}   & \,=\,  2^{\,1/2} \,\,|\,\CZ_{\,\CG}\,|
	 \,\equiv \,2^{\,1/2}\,|\,\CG \,\varPhi\,|~.
\label{m}
\\	 
 \CZ_{\,\CG}     &   \,=\,   g \cdot \AD+q\cdot A \,\equiv \,\CG \,\varPhi~.
\label{Z}
	\end{align}
Here $\CG$ represents the magnetic and electric charges of the dyon, while 
$\Phi$ describes the scalar field and its dual
\begin{align}
 \CG\,=\,(\,g,\,q\,)~,\quad\quad
 \Phi\,=\,(  A_\text{\it D}, \, A)^{\,T}~.
\label{Phi}
\end{align}
The discrete chiral transformations represent a symmetry of the theory
\begin{align}
&A\,\rightarrow \,A^{\,\prime}\,=\,\exp(2i\gamma)\,A~,
\label{AgA}
\\
& A_\text{\it D}\,\rightarrow\,A^{\,\prime}_D\,=\,\exp(2i\gamma)\,\big(\, A_\text{\it D}-
(2\gamma/\pi)\,\hv A\,\big)~.
\label{ADct}
\end{align}
Here $\gamma=2\pi m/(4\hv)$, $m=0,1,\dots 4\hv-1$, while 
$\hv$ is the dual Coxeter number of the gauge algebra, which is related to the eigenvalue of the quadratic Casimir operator $C_2$ in the adjoint representation, $2\hv=C_2$.
The transformation of the dual field in (\ref{ADct}) complies with its behavior at weak coupling
\begin{equation}
 A_\text{\it D}\,\approx\,(2\pi)^{-1}\,i\,\hv\, A\,\ln (\,A^2/\Lambda^2)~.
\label{ADPT}
\end{equation}
Here $A^2=A\cdot A$.  
For simplicity it is presumed here and below that the logarithmic factor is large, $|\ln A^2/ \Lambda^2|\gg 1$, and that $A$ is not close to a wall of the Weyl chamber.
Eqs. (\ref{AgA}), (\ref{ADct}) show that the defining element of the chiral symmetry is
\begin{equation}
\Phi\,\rightarrow \,
\Phi^{\,\prime} \, = \,
\exp\left( \,\pi i/\hv \right) H\,\Phi~,
	\label{ADgamma}
\end{equation}
where $H$  is the following $2r\times 2r$ matrix
\begin{equation}
H\,=\,\begin{pmatrix}~ 1 & \!-1~ \\ ~0 & \,~\,1~ \end{pmatrix}~.	
\label{Mch}
\end{equation}
The Witten and Olive formula (\ref{m}) implies that the chiral transformation of the field $\Phi$ can be reinterpreted as the transformation of the dyon charge
\begin{equation}
\CG\,\rightarrow \CG^{\,\prime}\,=\,\CG\,H~,
\label{GG'}
\end{equation}
which means that $g^{\,\prime}=g-q$,  $q^{\,\prime}=q$. The variation of the magnetic charge here agrees with the Witten effect \cite{Witten:1979ey}. 

Suppose there is  a dyon with magnetic and electric charges given by $\CG$, which becomes
massless at some value of the field $\varPhi$, $m_\CG=\sqrt{2}\,|\CG \,\varPhi|=0$. Then the dyon whose charge $\CG^{\,\prime}$ is generated from $\CG$ by the chiral transformation (\ref{GG'}) is also necessarily massless provided  the field equals
\begin{equation}
\varPhi^{\,\prime\prime}=\exp(-i\pi/\hv)H^{-1}\varPhi~,
\label{phi''}
\end{equation}
because  $m_\CG^{\,\prime}=\sqrt{2}\,|\,\CG^{\,\prime} \,\varPhi^{\,\prime\prime}|=\sqrt{2}\,|\,\CG \,\varPhi| =0$.
One concludes that if a monopole with an arbitrary magnetic charge $\CG=(g,0)$ is massless, then any dyon with the charge
\begin{equation}
\CG^{\,\prime}\,=\,\CG H^m\,=\,(g, -mg)~
\label{gen-light}
\end{equation}
can be made massless as well. Ref. \cite{Kuchiev:2009ez} emphasized that each dyon with the charge $\CG=(\av_i,-m\av_i)$, $i=1,\dots r$, $m=0,\dots \hv-1$ can be made massless if the state of the theory is chosen properly.

Eqs. (\ref{g=ng})  show that we can conveniently describe 
electric and magnetic charges in the basis of simple coroots  $\av_i$, $i=1,\dots r$ of the Cartan algebra. It is natural therefore to call $r$ monopoles whose magnetic charges equal simple coroots, $g=\av_i$, as simple monopoles. Similarly the electric charges, which equal the simple coroots $q=\av_i$, will be referred to as simple electric charges.


It was explained in \cite{Seiberg:1994rs} that the theory satisfies the important condition of duality, which amounts to the following transformation of the fields
\begin{align}
&\Phi\rightarrow \Phi^\prime\,=\,\Omega\,\Phi~,
\label{dual}
\\
&\Omega\,=\,\begin{pmatrix} ~~0 & 1~\\
													   -1 & 0~
													\end{pmatrix}~,
													\label{Omega}
\end{align}
that keeps the description of the theory intact.

Following \cite{Kuchiev:2009ez} we will describe the scalar and dual fields  
using the basis of fundamental weights $\omega_i$ of the Cartan algebra
\begin{equation}
A\,~=\,~\sum_{i=1}^r \,A_i~\omega_i~,
\quad\quad\quad\quad
 A_\text{\it D}\,=\,\sum_{i=1}^r\,A_{D,\,i}~\omega_i~,
\label{ADi}
\end{equation}
where $A_i$ and $A_{D,\,i}$ are the expansion coefficients.
This basis is convenient since the central charges (\ref{Z}) of dyons in this representation have a very clear form
\begin{equation}
\CZ_{\,\CG}   \,=\,  \sum_{i=1}^r\,\big(\, n^{(g)}_i \, A_{D,\,i} +n^{(q)}_i\, A_i\big) ~.
\label{Znn}
\end{equation}
Here integers $n^{(g)}_i$ and $n^{(q)}_i$ define the magnetic $g$ and electric $q$ charges in $\CG=(g,q)$ via  Eqs. (\ref{g=ng}).  
Deriving Eq. (\ref{Znn})  the known orthogonal condition 
\begin{equation}
\av_i\cdot\omega_j\,=\,\delta_{ij}
\label{avo}
\end{equation}
was used. From (\ref{Znn}) we see that each $A_i$ equals the central charge related to the simple electric charge  $q=\av_i$; similarly each $A_{D,\,i}$ equals the
central charge of the monopole with magnetic charge $g=\av_i$
\begin{equation}
\CZ_i\equiv \CZ_{\,(0,\av_i)}=A_i~,\quad\quad \CZ_{D,\,i}\equiv\CZ_{\,(\av_i,0)}=A_{D,\,i}~.
\label{ZiZDi}
\end{equation} 
Thus $A_i$ and $A_{D,\,i}$ can be viewed as either expansion coefficients for the scalar and dual fields in the basis of fundamental weights, or central charges for simple 
electric charges or simple monopoles respectively.  This fact and the clear form of Eq. (\ref{Znn}) inspire one to employ the sets of $A_i$ and $A_{D,\,i}$ 
as arguments of the superpotential $\CF(A)$ and its dual $\CF_D(A_D)$
\begin{align}
\CF(A)\,=\,\CF(A_1,\,\dots \,A_r)~,
\quad\quad\quad
\CF_D(A_D)\,=\,\CF_D(A_{D,\,1},\,\dots\, A_{D,\,r})~.
\label{FDAD}
\end{align}
For our purposes it suffices to restrict discussion below to those values of $A$, which belong to the main Weyl chamber. This means that all $A_i$ are presumed real positive, $A_i>0$. The idea is that whence the theory is formulated in this region, it can be continued for arbitrary values of $A$ using the analytical continuation.

\section{Central charges and discrete symmetry}

Eqs.(\ref{g=ng}) state that the electric and magnetic charges are pinned to vertexes 
of the lattice of coroots ${\mathbb Q}^\vee$ of the Cartan algebra. 
The existence of this lattice should be compatible with the low-energy description of the theory. From general physical grounds one may expect that this compatibility can be achieved only if the theory complies with some particular discrete symmetry related to the lattice. (Electrons propagating in the crystal lattice may serve as an example.) 
 
The main idea is to use the central charges as a tool for searching for the discrete symmetry in question. Take the full set of all central charges
$\mathfrak{Z}$ available in the theory
\begin{equation}
\mathfrak{Z}= \{\CZ_{(g,\,q)},~g,q\in \mathbb{Q}^\vee\}~.
\label{ZZ}
\end{equation}  
If some transformation represents a symmetry of the system, then 
there are only two options. Either this transformation
keeps the full set $\mathfrak{Z}$ intact, or alternatively 
it changes this set in such a way that the same phase factor 
is added to all central charges in this set, 
$\mathfrak{Z}\rightarrow \mathfrak{Z}^{\,\prime}=
\{\CZ_{(g,\,q)}^{\,\prime}=e^{2i\gamma}\CZ_{(g,\,q)},~g,q\in \mathbb{Q}^\vee\}$.
For example, under the dual transformation (\ref{dual}) the set $\mathfrak{Z}$ remains intact, while the discrete chiral transformation in (\ref{AgA}), (\ref{ADct}) adds a phase factor $e^{2i\gamma}$ to all central charges.  Any other variation of the set $\mathfrak Z$, except for a common  phase factor, would change masses of some dyons in Eq.(\ref{m}) and therefore would contradict the fact that the transformation is a symmetry of the system. 

In the following discussion it suffices for us to consider only those transformations, which provide no additional phase factors to central charges, $\gamma=0$. 
Let us reiterate that in this case the invariance of the set $\mathfrak{Z}$ presents a necessary condition for a transformation to be a symmetry. Moreover, there is a good reason to believe that this is also a sufficient condition, which guarantees that the chosen transformation is definitely a symmetry. In other words, if the state of the supersymmetric $\CN=2$ gauge theory is transformed in such a way that the set of central charges $\mathfrak{Z}$ is kept intact, then the related transformation is a symmetry. 
To support this claim remember that the set of central charges completely defines the $\CN=2$ supersymmetric algebra. As a result all vital properties of the system can be formulated in terms of central charges of dyons. By keeping the set of central charges intact one keeps the 
the system properties unchanged. Thus the invariance of $\mathfrak{Z}$ guarantees the presence of the symmetry. (We will see below in Section \ref{model} that this condition is reproduced in the model of \cite{Kuchiev:2009ez}, which complies with all other known conditions required by the theory.)

Consider a transformation  $A\rightarrow A^{\,\prime}$
of the scalar field, which amounts to a permutation of the set $\{A_i,\,i=1,\dots r\}$
and hence can be presented as follows
\begin{equation}
A_i^{\,\prime}\,=\,\sum_{j=1}^r\mathcal{P}_{ij} \,A_j~.
\label{P}
\end{equation}
Here $\mathcal{P}$ is the $r\times r$ matrix representing the permutation of $r$ objects. 
To clarify notation let us write the permutation conventionally, using the integer function $p(j)$ of an integer argument $j$, which satisfies conditions $1\le  p(i)\le r$ and $p(i)\ne p (j)$ if $i\ne j$, $i,j=1,\dots r$.
Then the matrix $\mathcal P$ reads $\mathcal P_{ij}=\delta_{\,i,\, p(j)}\,$.
Alternatively this matrix can be specified as an orthogonal matrix, $\mathcal{P}^T\mathcal{P}=1$, which coefficients take one of the two values,
zero or unity, $\mathcal{P}_{ij}=0,1$.
For such matrices we will use notation $\mathcal{P}\in S_r$, where $S_r$ is the group of all 
available $r!$ permutations.
The shortcut matrix notation $A^{\,\prime}=\mathcal P A$ is used below to present (\ref{P}) and similar relations.

To clarify the reason for the choice $\mathcal{P}\in S_r$ in (\ref{P}) let us allow ourselves for a moment to make an additional assumption (which is justifies below), that this transformation of the scalar field (\ref{P}) causes a similar transformation for the dual field, $A_D\rightarrow {A^{\,\prime}_D}$, where
\begin{equation}
A_{D,\,i}^{\,\prime}\,=\,\sum_{j=1}^r\mathcal{P}_{ij}\, A_{D,\,j}~,
\label{Pstrong}
\end{equation}
and $\mathcal{P}$ is the same as in (\ref{P}). Then we can use  Eq. (\ref{Znn}) to show that under the transformation $(A_D, A)\rightarrow (A^{\,\prime}_D, A^{\,\prime})$ defined in (\ref{P}), (\ref{Pstrong}) the set of central charges $\mathfrak Z$  in (\ref{ZZ}) remains invariant. 
To see this point take the electric charge $q=\sum_i n^{(q)}_i\av_i$. The corresponding central charge reads
$\CZ_{(0,q)}=\sum_i n^{(q)}_i A_i$. 
Under the transformation $A\rightarrow A^{\,\prime}$
from (\ref{P}) the central charge is transformed as follows
$\CZ_{(0,q)}\rightarrow 
\CZ^{\,\prime}_{(0,q)} =\sum_{ij} n^{(q)}_i \mathcal P_{ij} A_j=\sum_{i} n^{(q)\,\prime}_i A_i$,
where $n^{\,\prime}_i=\sum_{j} \mathcal P^{\,T}_{ij} \,n_j$. Since  $\mathcal P^T \in S_r$, $n^{(q)\,\prime}_i\in \mathbb Z$. Therefore we find $\CG^{\,\prime}_{(0,q)}=
\CZ_{(0,q^{\prime})}$, where $q^{\,\prime}=\sum_i n^{(q)\,\prime}_i\av_i$ is a new electric charge. In other words, the influence of the transformation $A\rightarrow A^{\,\prime}$ from (\ref{P}) on the central charge can be redefined using the transformation for the electric charge $q\rightarrow q^{\,\prime}$. Similarly the
transformation $A_D\rightarrow A^{\,\prime}_D$ from (\ref{Pstrong}) can 
be reinterpreted as the variation of magnetic charges $g\rightarrow g^{\,\prime}$.
In both electric and magnetic cases this reinterpretation of the transformation reads
\begin{equation}
 n^{(q)\,\prime}_i\,=\,\sum_{j=1}^r \mathcal{P}^T_{ij}\,n^{(q)}_j~,
 \quad\quad
 n^{(g)\,\prime}_i\,=\,\sum_{j=1}^r \mathcal{P}^T_{ij}\,n^{(g)}_j~.
 \label{nmag}
\end{equation}
From this perspective the considered transformation amounts to the permutation of all available in the theory electric and magnetic charges by the matrix $\mathcal P^T\in S_r$, while $A$ and $A_D$ remain intact. Obviously under this permutation no charge is lost and none is created. Hence, the full set of all available dyons remains intact, which in turn means that the set of all central charges $\mathfrak Z$ remains intact as well. 

Using the invariance of $\mathfrak Z$ under $\mathcal{P}\in S_r$  as the criterion which justifies the presence of the symmetry we conclude that the group $S_r$ represents a symmetry group of the theory. We have to remember though that so far Eq.(\ref{Pstrong}), which was employed in this argument, was only assumed to be valid. 

To verify (\ref{Pstrong}) start from the weak coupling region. Take Eq.(\ref{ADPT}) and apply the transformation (\ref{P}) to the scalar field $A$ on its right hand side.  
Observe then that a similar permutation inevitably takes place with the dual filed $A_D$ on the left hand side. Making this claim one remembers that the logarithm in (\ref{ADPT}) is large, $A$ not close to walls of the Weyl chamber, and hence $\ln A^2\approx\ln A^{\,\prime\,2}$.  
Thus, for weak coupling the permutation of $A_i$ in (\ref{P}) is necessarily accompanied by 
the permutation of $A_{D,\,i}$ in (\ref{Pstrong}), and the resulting pair of transformations (\ref{P}) and (\ref{Pstrong}) keep $\mathfrak Z$ invariant.

Remember that the dual field $A_D$ can be considered a function of the scalar field $A$ expressed via derivatives of the superpotential $\CF(A)$ in Eq.(\ref{FDAD}).  The superpotential is holomorphic \cite{Seiberg:1988ur}, which implies that there exist $r$ holomorphic functions representing $A_{D,\,i}$ via $r$ arguments $A_i$. 
Therefore we can presume that at weak coupling the considered discrete permutation is fulfilled via a smooth variation of both fields. The  starting point for the function, which 
describes this variation is $(A_D,A)$, its final point is $(A^{\,\prime}_D, A^{\,\prime})$ while in the intermediate state both fields exhibit holomorphic variation.

Since the compliance of (\ref{Pstrong}) with (\ref{P}) is established via an analytical continuation one can develop the argument by analytically continuing
the fields from the weak coupling region into the area of an arbitrary coupling. The point is that since $\mathcal P\in S_r$ is a discrete transformation with integer matrix elements. Therefore it cannot be changed by an analytical continuation.
As a result we conclude that the necessary property, 
the compliance of (\ref{Pstrong}) with (\ref{P}) remains valid at arbitrary coupling.
The argument can be reversed. Take $A$ as a function
of $A_D$, which is expressed via the derivatives of the dual superpotential $\CF_D(A_D)$,
express $A_i$ as $r$ analytical functions of arguments $A_{D,\,i}$ and then
state that (\ref{P}) follows from (\ref{Pstrong}).

We conclude that the theory is invariant under transformations $\mathcal P\in S_r$ described in  Eqs.(\ref{P}) and  (\ref{Pstrong}). These two equations are mutually related, a transformation of $A_i$ necessitates a similar transformation for $A_{D,\,i}$, and vice versa.  The two equations can be conveniently presented as follows
\begin{equation}
\Phi\,\rightarrow \,\Phi^{\,\prime} \, = \,P\,\Phi~,
\label{PPhi}
\quad\quad
P\,=\,\begin{pmatrix}~ \mathcal{P} & 0~ \\ 0 & \mathcal{P}~ \end{pmatrix}~,
\end{equation}
where $\mathcal{P}\in S_r$ and the previously mentioned shortcut matrix notation presents transformations of $A$ and $A_{D}$, for example $A^{\,\prime}=\mathcal PA$ means  $A^{\,\prime}_i=\sum_j {\cal P}_{ij}A_j$. 

Since we know that $S_r$ is a symmetry we need to consider its representations. They can be conveniently described in terms of the symmetry properties
of the superpotential, which uniquely describes the system. We need thus to specify in which representation of $S_r$ the superpotential $\CF(A)$ and its dual $\CF_D(A_D)$ belong. 
Clearly the representations in question should be  irreducible. Otherwise the splitting of the system into subsystems looks inevitable, which would contradict the fact that the gauge group is simple.
Moreover, it is easy to see that the superpotential and its dual belong to the trivial representation of $S_r$, which means that they both are symmetric functions of
their arguments $A_i$ and $A_{D,\,i}$
\begin{equation}
\CF(A_1^{\prime},\dots A_r^{\prime})=\CF(A_1,\dots A_r),
\label{Fsym}
\quad~
\CF(A_{D,\,1}^{\prime},\dots A_{D,\,r}^{\prime})=\CF(A_{D,\,1},\dots A_{D,\,r}).
\end{equation}
Here $A^{\,\prime}$ and $A^{\,\prime}_{D}$ are obtained by applying  $\mathcal P \in S_r$
to $A$ and $A_{D}$ as described  in Eqs. (\ref{P}), (\ref{Pstrong}), or (\ref{PPhi}). 
To verify (\ref{Fsym})  assume the opposite, that $\CF(A)$ is in any nontrivial irreducible representation  of $S_r$ described by some Young scheme. Then it is a function antisymmetric under a transposition of any arguments $A_i$, $A_j$, which belong to the same column of the Young scheme. Therefore $\CF(A)$  turns zero identically when $A_i=A_j$, which clearly contradicts its behavior at weak coupling. 
A similar argument is valid for $\CF_D(A_{D})$ since in this case we can rely on its behavior at strong coupling. Thus  the symmetry conditions (\ref{Fsym}) provide the only option for representing $S_r$.

It is tempting to extend the found $S_r$ symmetry to a larger symmetry group.
A seemingly good option present the set of all matrices $\mathcal P$, which are orthogonal and have integer valued coefficients. If such $\mathcal{P}$ is diagonal, then it may have negative diagonal matrix elements $\mathcal{P}_{ii}=- 1$, which describe inversions of the corresponding of $A_i$, $A_i\rightarrow A^{\,\prime}_i=-A_i$.  
Combining such inversions with permutations from $S_r$ one recovers the full set of matrices $\{ \mathcal{P}\in O(r),\,\mathcal{P}_{ij}\in \mathbb Z\,\}$, which represents the full symmetry group of the  $r$-cube
(it includes $2^{\,r} r!$ elements, is called the hyperoctahedral group, and in the Coxeter notation is of type $C_r$, here $C_r$ should not be confused with the gauge group $G$ of the considered theory). Thus the temptation is to extend the previously considered arguments proclaiming that the symmetry group of the $r$-cube represents the symmetry of the supersymmetric $\CN=2$ gauge theory.

There is a subtlety though, which makes such straightforward conclusion premature. 
Developing the argument above we had in mind that the transformation $(A_D,A)\rightarrow (A^{\,\prime}_D,A^{\,\prime})$ 
can be fulfilled via a holomorphic set of functions. 
It is reasonable to expect that this is possible when a transposition  $A_i\rightarrow A^{\,\prime}_i=A_j$, $A_j\rightarrow A^{\,\prime}_j=A_i$ is considered, which 
justifies the argument for any permutation from $S_r$.
However, an attempt to consider the inversion of some of $A_i$, $A_i\rightarrow A^{\,\prime}_i=-A_i$, requires the analytical continuation
through the wall of the Weyl chamber, which is specified by the condition $A_i=0$. This condition indicates that the central charge and mass, which correspond to the electric charge
$q=\av_i$ are zero on the wall, see Eq.(\ref{Znn}). Hence one needs to expect  that 
on the wall $A_i=0$ of the Weyl chamber there exists a  singularity, and that the crossing of the wall should be considered carefully. The way to account for the necessary crossing was discussed in Ref. \cite{Kuchiev:2009ez}, see Section 9 there. It would be interesting to consider implications when it is combined with the results of the present work, but we 
postpone discussion of this point restricting our attention in the present work
to the $S_r$ symmetry, which represents a rewarding topic by itself.

The starting point of our arguments was the Dirac-Schwinger-Zwanziger quantization condition, which forces the electric and magnetic charges to form the lattice $\mathbb Q^\vee$.
However, the found symmetry group $S_r$ is a subgroup of the full symmetry group $C_r$ of the $r$-cube, $S_r\subset C_r$. Thus the detailed geometrical structure of
$\mathbb Q^\vee$, which is incorporated into the Cartan matrix 
(that describes relative lengths and angles between the coroots $\av_i$), 
does not show  itself in the found symmetry group. 

The symmetry group of the $r$-cube is higher than the symmetry group of the lattice $\mathbb Q^\vee$. In accord with this fact the found symmetry $S_r$ is higher than one may have expected. The reason for this symmetry ``enhancement" stems from the fact that 
the lattice of electric and magnetic charges $\mathbb Q^\vee$ manifests itself in the theory via the set of central charges $\mathfrak Z$.
Due to this reason the transformation of the scalar and dual fields in Eq.(\ref{P}), (\ref{Pstrong}) can be rewritten in terms of transformations of integers $n^{(q)}_i$ and $n^{(g)}_i$, which describe the electric and magnetic charges of dyons, see Eqs.(\ref{nmag}). The sets of these integers can be looked at as vertexes of the $r$-cubic lattice, which 
turns out to be the only lattice that matters. Thus the symmetry $S_r$, which is related to the $r$-cube, becomes relevant.

To simplify discussion it is convenient to interpret Eqs.(\ref{Fsym}) as
a symmetry of the system under a simultaneous permutation of $r$ simple electric charges and $r$ simple monopoles.

\section{Duality,  permutations and central charges}

The duality for the supersymmetric $\CN=2$ theory introduced in
\cite{Seiberg:1994rs} for the SU(2) gauge symmetry can be described 
via a Legendre-type transformation $\CF(A)\rightarrow \CF_D(A_D)=AA_D-\CF(A)$, in which the product $\Xi= AA_D$ plays the role of a generating function for this transform. 
In order to extend the dual conditions to the theory with an arbitrary gauge symmetry one
can follow a similar approach, considering the Legendre-type transform, in which
the generating function $\Xi$ is a bilinear form of the fields $A$ and $A_D$. 
Generally speaking for $r>1$ there exists a variety of different bilinear forms, 
and we need to find the right one. 

This task can be resolved using the symmetry conditions (\ref{Fsym})  which allow only two options,  one of which, $\Xi=\sum_{i=1}^r A_iA_{D,\,i}$, 
is suitable for our needs. Another one, $\Xi^{\,\prime}=\sum_{i=1}^r \sum_{j=1}^r A_i A_{D,\,j}$, should be rejected. The reason for this rebuff becomes clear at weak coupling, where one expects the simple electric charges to behave independently. In contrast to this expectation the bilinear $\Xi^{\,\prime}$ strongly mixes the related variables and hence should be rejected.

Consequently we present the dual transformation for an arbitrary gauge group as follows
\begin{equation}
\CF_D(A_D)=\sum_{i=1}^{r} A_iA_{D,\,i}-\CF(A)~. 
\label{Fdual}
\end{equation}
Differentiating it over $A_i$ and $A_{D,\,i}$ and
using the fact that $\CF(A_D)$ is $A$-independent, while 
$\CF(A)$ is $A_D$-independent, we find
\begin{equation}
A_{D,\,i}\,=\frac{\partial \CF}{\partial A_i}~,
\quad\quad\quad
A_{i}~=-\frac{\partial \CF_D}{\partial A_{D,\,i}}~.
\label{AD=dF/dA}
\end{equation}
These transparent relations demonstrate again the advantage of 
expanding $A$ and $A_D$  via the fundamental weights in (\ref{ADi}),  
or presenting these fields via the central charges for electric charges and monopoles, see (\ref{ZiZDi}).
The $\tau$-matrix of the coupling constants in this basis reads
\begin{equation}
\tau_{ij}(A)\,=\,
\frac{\partial A_{D,\,i} }{\partial A_{j} }\,=\,
\frac{\partial^2 \CF(A)}{\partial A_{i}\,\partial A_{j}}~.
\label{tau}
\end{equation}
The invariance  of the superpotential 
under $S_r$ in (\ref{Fsym}) implies the transformation of the $\tau$ matrix.
If $A^{\,\prime}=\mathcal P A$, see (\ref{P}), then
\begin{equation}
\tau(A^{\,\prime})\,=\,{\cal P}\,\tau(A)\,{\cal P}^{\,T}~,
\label{Ptau}
\end{equation}

\section{Weak coupling and $\mathbf S_r$}
\label{Weak coupling}
It is instructive to apply the formalism discussed above to the region of weak coupling. 
Combining  Eq.(\ref{tau}) with (\ref{ADPT}) we find
\begin{equation}
\tau_{ij}\,\approx\,\delta_{ij}\,\frac{i}{2\pi}\,\hv \,\ln\frac{A^2}{\Lambda^2}
\,\approx\,\delta_{ij}\,\frac{i}{2\pi}\,\hv \,\ln\frac{\sum_i A_i^2}{\Lambda^2}~.
\label{tauPTh}
\end{equation}
Similarly to Eq.(\ref{ADPT}) the logarithm here is presumed large $|\ln A^2/\Lambda^2| \gg 1$ and $A$ is not close to a wall of the Weyl chamber. Using this fact we replaced 
the argument in the logarithmic function
$\ln(A^2/\Lambda^2)\approx\ln\big(\sum_{i=1}^r A^2_i/\Lambda^2\big)$.

Eq.(\ref{tauPTh}) shows that the $\tau$-matrix is diagonal in the basis of the fundamental weights. As a corollary we find that in any other basis conventional for the Cartan algebra, orthogonal basis, basis of simple roots or simple coroots, the $\tau$-matrix is definitely non-diagonal. To illustrate this fact consider the Cartan algebra in the basis of orthonormal vectors $\varepsilon_s$,  $\varepsilon_s\cdot \varepsilon_t=\delta_{st}$, $s,t=1,\dots r$. 
Expanding the field $A$ in this basis we write
\begin{equation}
A\,=\,\sum_{s=1}^{r} A^{\,\perp}_s\,\varepsilon_s~,
\label{perp0}
\end{equation}
where the superscript $\perp$ distinguishes the coefficients of this expansion $A^{\,\perp}_s$ (from the expansion coefficients $A_i$ in the basis of fundamental weights in Eq.(\ref{ADi})).  Transforming (\ref{tauPTh}) to the orthogonal basis $\tau\rightarrow \tau^\perp$ we find 
\begin{equation}
\tau^{\perp}_{st}\,=\,\frac{\partial^2 \CF(A)}{\partial A^{\,\perp}_{s}\,\partial A^{\,\perp}_{t}}
\,\approx\,{\cal Q}^{-1}_{st}~\frac{i}{2\pi}\,\ln\frac{\sum_i A_i^2}{\Lambda^2}~.
\label{tauPThOrt}
\end{equation}
Here ${\cal Q}^{-1}$ is the inverse of
the $r\times r$ matrix ${\cal Q}$, which represents the well known quadratic form  for the Cartan algebra
\begin{equation}
{\cal Q}_{\,ij}=\omega_i\cdot\omega_j~.
\label{Y}
\end{equation}
To verify Eq.(\ref{tauPThOrt}) one
uses Eq.(\ref{avo})  to show that $\partial A_i/\partial A^{\,\perp}_s =(\av_i\cdot \varepsilon_s)$ and applies also the equality
\begin{equation}
{ \cal Q }^{-1}_{st}\,=\,
\sum_{i=1}^r \,(\av_i\cdot \varepsilon_s)\,(\av_i\cdot \varepsilon_t)~.
\label{Q-1}
\end{equation}
To validate this last relation one takes two $r\times r$ matrices  
$\omega$ and $\av$ with matrix elements 
$\omega_{is}=\omega_i\cdot \varepsilon_s$ and $\av_{is}=\av_i\cdot\varepsilon_s$, 
rewrites Eqs. (\ref{avo}), (\ref{Y}) and (\ref{Q-1}) in the matrix form, 
$\av \,\omega^T=1$, ${\cal Q}=\omega \,\omega^T$, and ${\cal Q}^{-1}=(\av)^T\av$,
and immediately observes that the third among them follows from the first two,
in other words that (\ref{Q-1}) follows from (\ref{avo}) and (\ref{Y}). 
Remember now that for any simple gauge group the quadratic form ${\cal Q}$ is non-diagonal, which makes  $\tau^{\,\perp}$ in Eq.(\ref{tauPThOrt}) non-diagonal as well.

Combining  Eq.(\ref{tauPTh}) with (\ref{tau}) we derive the following expression for the  superpotential at weak coupling 
\begin{equation}
\CF(A)\,\approx\,\frac{i}{4\pi}\,\hv\,\sum_{i=1}^r \,A_i^2\,\ln\frac{\sum_j A_j^2}{\Lambda^2}~.
\label{Fweak}
\end{equation}
Clearly it is explicitly invariant under $\mathcal{P}\in S_r$ from (\ref{P}), which agrees with our expectations (\ref{Fsym}).

Compare Eq.(\ref{Fweak}) with the well known expression for the superpotential at weak coupling, which was extensively used in the literature previously 
\begin{equation}
\CF_0(A)\,\approx\,\frac{i}{8\pi}\sum_\alpha \,(A\cdot \alpha)^2\ln\frac{A^2}{\Lambda^2}
\,\approx\, \frac{i}{4\pi} \,\hv\,A^2\ln\frac{A^2}{\Lambda^2}~.
\label{Ftrad}
\end{equation}
The summation in the middle expression here runs over all roots.
To derive the final result one takes into account the identity
\begin{equation}
\sum_\alpha\alpha \otimes \alpha =2\hv~,
\label{aa2h}
\end{equation}
and presumes once again that since the logarithm is large and $A$ is not close to a wall of the Weyl chamber the logarithmic factor is a smooth function of $A$, and as such can be taken out of summation.

Note  an important distinction separating $\CF(A)$ from $\CF_0(A)$ in Eqs. (\ref{Fweak}) and (\ref{Ftrad}) respectively. 
The superpotential $\CF(A)$ incorporates the factor $\sum_{i=1}^r A_i^2$, which is symmetric under the permutations from $S_r$ defined in (\ref{Fsym}). 
In contrast $\CF_0(A)$ possesses instead a factor $A^2=\sum_{s=1}^r (A^\perp_s)^2$, which shows no signs of the necessary symmetry $S_r$. Hence, in line with the arguments of the present work, it should be considered misleading, see also
Appendix \ref{subtlety}.

\section{Alignment of scalar fields along Weyl vector}
\label{Degeneracy-Weyl}

In general there exist no specific reasons that would force the two vectors $A$ and $A_D$ to be collinear. There is though an important exception. If one of these two vectors, $A$ or $A_D$, is aligned along the Weyl vector of the Cartan algebra $\rho$, then another one is necessarily aligned along the same direction, which makes all three of them
collinear. This property can be represented as the following statement
\begin{equation}
A\,=k\,\rho \quad \Longleftrightarrow \quad A_D \,=\,k_D\,\rho~,
\label{Weyl}
\end{equation}
where $k$ and $k_D$ are numbers.
We will see below that Eq.(\ref{Weyl}), which will be referred to as the Weyl vector alignment, has interesting physical consequences.

To verify that the alignment along the Weyl vector  (\ref{Weyl}) takes place recall that the Weyl vector equals the semi-sum of positive roots, or equivalently the sum of the fundamental weights $\omega_i$
\begin{equation}
\rho\,=\,\frac{1}{2}\,\sum_{\alpha>0}\,\alpha\,=\,\sum_{i=1}^r \,\omega_i~.
\label{ro}
\end{equation}
Consequently, if $A$ is aligned along the Weyl vector, $A=k\,\rho$ where $k$ is a number, then all expansion coefficients of $A$ in the basis of simple weights in Eq.(\ref{ADi}) are identical, $A_i=k$, $i=1,\dots r$. From Eq.(\ref{AD=dF/dA}) we see then
that $A_{D,\,i}=\partial_i \CF(k,\,\dots k)$, where $\partial_i \equiv \partial/\partial A_{D,i}$. Remembering now that $\CF(A)$ is symmetrical under the permutations of its arguments
(\ref{Fsym}), we conclude that all $A_{D,\,i}$ are same, $A_{D,\,i}= k_D$ where $k_D=
\partial_1\CF(k,\,\dots k)$ is a number.
The identity of the expansion coefficients $A_{D,\,i}$ in Eq.(\ref{ADi}) means that
the vector $A_D$ is parallel to $\rho$ defined in (\ref{ro}), 
$A_D=k_D\rho$. Thus, we verified that the relation $A=k\,\rho$ results in $A_D=k_D\,\rho$. Similarly we investigate the opposite direction finding then
that $A_D=k_D\,\rho$ imposes $A=k\,\rho$, which
proves (\ref{Weyl}). 
Note that this discussion was based entirely on the single property of the superpotential, 
its symmetry under permutations in Eqs. (\ref{Fsym}).

As was mentioned, the alignment of $A_D$ along the Weyl vector means that all
its expansion coefficients $A_{D,\,i}$ are same, $A_{D,\,i}=k_D$. In turn, according to  (\ref{Znn}) this indicates that the central charges of all $r$ simple monopoles are 
degenerate, have the central charge $\CZ_{(\av_i,0)}=k_D$. 
Call this situation a complete degeneracy of central charges of monopoles.
A central charge and mass for an arbitrary monopole in this case reads
\begin{align}
&\CZ_{(g,0)}\,=\,k_D\,\sum_{i=1}^r  n^{(g)}_i~,
\label{Zg}
\\
&m_{(g,0)}\,=\,\sqrt2 ~\big|k_D\,\sum_{i=1}^r  n^{(g)}_i\big|~.
\label{mg}
\end{align}
where integers $n^{(g)}$ define the magnetic charge $g$ via Eq.(\ref{g=ng}).
Thus all central charges of monopoles are defined via the only complex constant
$k_D$, while the mass of any monopole is expressed via its absolute value.
In particular, masses of all $r$ simple monopoles are degenerate.

Similarly we find that the alignment of the scalar field along the Weyl vector
means that the central charges of $r$ simple electric charges are same
$\CZ_{(0,\av_i)}=A_i=k$, which signals their complete degeneracy.
The central charge and mass of any electric charge in this case read
\begin{align}
&\CZ_{(0,q)}\,=\,k\,\sum_{i=1}^r  n^{(q)}_i~,
\label{Zq}
\\
&m_{(0,q)}\,=\,\sqrt2 ~\big|k\,\sum_{i=1}^r  n^{(q)}_i\big|~,
\label{mq}
\end{align}
being defined by the constant $k$ and its absolute value respectively.
Eq.(\ref{mq}) means, in particular, that masses of all $r$ simple monopoles are degenerate.

The alignment of $A$ and $A_D$ along the Weyl vector specified in (\ref{Weyl}) can be reformulated now in terms of degeneracy of central charges. From Eqs. (\ref{Weyl}), (\ref{Zg}) and (\ref{Zq}) we conclude that if the central charges of monopoles are completely degenerate (central charges of $r$ simple monopoles are equal), then the central charges for electric charges are necessarily completely degenerate as well
(central charges for $r$ simple electric charges are same). Vice versa, the complete degeneracy of central charges among electric charges implies complete degeneracy 
of central charges for monopoles.

This observation looks attractive 
and blends well within the general idea of duality. 
Under the dual transformation the simple monopoles
and simple electric charges should switch places, 
similarly the dual and scalar field switch their places as well. 
It seems desirable that this switch 
transforms a possible degeneracy, which may exist in the spectrum of monopoles,
into a similar degeneracy in the spectrum of electric charges.
It is gratifying therefore that the alignment of $A$ and $A_D$ along the Weyl vector ensures that the complete degeneracy is invariant under the duality transformation.

This discussion  can be extended to cover the case of dyons with charges
$ \CG\,=\,(\,g,\,-m g\,)$.
Relying on the Weyl vector alignment (\ref{Weyl}) one finds 
that the central charges of $r$ dyons 
with charges $ \CG\,=\,(\,\av_i,\,-m \av_i\,)$, are all degenerate
$\CZ_{\,\CG}\,=\,k-mk_D$.
Consequently the masses of these $r$ dyons are also degenerate, while the mass of any  
dyon with the charge $\CG\,=\,(\,\av_i,\,-m \av_i\,)$, where $g$ is arbitrary, reads $m_\CG=2^{1/2}|(k-mk_D)\sum_i n^{(g)}_i|$.

\section{Zero dual field, zero monopole masses}

A very interesting implication of the Weyl vector alignment (\ref{Weyl})  
arises in the limit $A_\text{\it D}\rightarrow 0$. 
Take this limit in such a way that $A_\text{\it D}$ remains aligned along the Weyl vector, $A_\text{\it D}\parallel \rho $. 
Then Eq.(\ref{Weyl}) ensures that in the process $A$ also remains 
parallel to the Weyl vector. Consequently we find that the node of the dual field $A_\text{\it D}=0$ takes place when the scalar field is aligned along the Weyl vector.
In accord with Eq.(\ref{m}) the condition $A_\text{\it D}=0$ is equivalent to the statement that all monopoles are massless
$m_g\,=\,0$,
where $g$ is an arbitrary magnetic charge.
We conclude that the Weyl vector alignment (\ref{Weyl}) 
implies that monopoles become massless at some particular value of $A$, which is aligned along the Weyl vector. For future references let us present this important statement as follows
\begin{equation}
A_\text{\it D}=0\quad\Longrightarrow \quad A\,=\, c \,\Lambda\, \rho~.
\label{corollary}
\end{equation}
Here $\Lambda$ is a cutoff parameter of the theory, which is written
on the basis of simple dimensional counting, while $c$ is a number.
Addressing the issue below with the help of the model of \cite{Kuchiev:2009ez}
we calculate $c$ explicitly in Eqs. (\ref{Af}), (\ref{k}). 

The point where the dual field turns zero plays an exceptionally important role in the theory. First of all at this point all monopoles become massless.
Secondly, as was pointed out in \cite{Seiberg:1994rs}, at this point the $\CN=2$ supersymmetric gauge theory can be broken explicitly down to $\CN=1$ supersymmetric gauge theory, and hence it can be considered a point of $\CN=1,2$ transition. It is rewarding therefore that an analysis based entirely on general properties of the theory, involving no model calculations, specifies the related value of the field $A$ in (\ref{corollary}) in such detail. 

The alignment of $A$ along the Weyl vector established in (\ref{corollary}) indicates  that at the point where $A_D=0$ the spectrum of electric charges turns completely degenerate, i. e. central charges that correspond to $r$ simple electric charges are all identical  $\CZ_i=c\,\Lambda$, $i=1,\dots r$; by the same token central charges and masses 
of all electric charges are specified by Eqs.(\ref{Zq}) and (\ref{mq}) in which $k=c\,\Lambda$.

The discrete chiral invariance allows us to generalize this result for a massless set of dyons with charges $\CG_\text{dyon}=(g,-mg)$, where $m$ is fixed and $g$ is an arbitrary positive coroot. To this end rewrite (\ref{corollary}) in the following convenient form
\begin{equation}
\Phi_\text{mon}\,=\,k_\text{mon}\,\Lambda\begin{pmatrix} 0 \\ \rho \end{pmatrix}~,
\label{Pho0}
\end{equation}
where $k_\text{mon}$ is a constant.
Then apply Eq.(\ref{phi''}) to  find out that the field, which makes massless the set of dyons 
with charges $\CG_\text{dyon}=(g,-mg)$ (where $m$ is fixed) reads
\begin{equation}
\Phi_\text{dyon}\,=\,
\exp\,\left(\!-\frac{m\pi i}{\hv}\,\right)\,k_\text{mon}\,\Lambda \,\begin{pmatrix} m\,\rho\\ \rho\end{pmatrix}~,
\label{Phim=}
\end{equation}
which in explicit notation gives
$A=\exp(-{m\pi i}/{\hv})\,k_\text{mon}\,\Lambda \,\rho$ and
$A_\text{\it D}=m\,A$.

\section{Symmetry $S_r$ in model approach}
\label{model}
Recall the main features of the recently suggested model \cite{Kuchiev:2009ez}, 
which generalizes the Seiberg-Witten approach \cite{Seiberg:1994rs,Seiberg:1994aj} 
to the supersymmetric $\CN=2$ gauge theory to describe a theory with arbitrary gauge group.
The coefficients $A_i$ and $A_{D,\,i}$ 
in the expansion of the fields in Eqs.(\ref{ADi}) in this model
are presented via the Abelian integrals
\begin{equation}
A_i\,~=\,~\frac{1}{2\pi i}~\oint_{C_i}~d\lambda~,
\quad\quad\quad
A_{D,\,i}\,=\,\frac{1}{2\pi i}\,\oint_{C_{D,i}}\!\!d\lambda~,
\label{Adl}
\end{equation}
where
\begin{equation}
d\lambda\,=\,\sqrt{z}~\frac{P^{\,\prime}(z) }{Y(z)}\,dz~.
\label{dl}
\end{equation}
Here $Y(z)$ is the hyperelliptic curve defined as follows
\begin{align}
& Y^2(z)\,=\,P^2(z)-Q^2~,
\label{YPQ}
\\
& P(z)\,=\,\prod_{i=1}^r\,(z-a_i^2)~,
\quad\quad
 Q\,=\,[\,a\,]^{2r-\hv}\,\Lambda^{\hv}~,
\label{P(z)}
\\
&[\, a\,]\,=\,\Big(~\frac{1}{r}~\sum_{i=1}^r\,(a_i)^{\,2r}\,\Big)^{1/(2r)}~.
\label{f2}
\end{align}
The nodes $a_i^2$ of the polynomial $P(z)$
at weak coupling describe  the classical value $a$ of the scalar field
\begin{equation}
a=\sum_{i=1}^r \,a_i\,\omega_i~,
\label{p-th}
\end{equation}
which approximates the proper scalar field $A$, $A\approx a$ when $\Lambda\rightarrow 0$.
The factor $\Lambda$ in Eq.(\ref{P(z)}) is the cutoff parameter. 

The differential $d\lambda$ possesses singularities at $z=0$ and $z=\infty$  from the factor $\sqrt{z}$, and singularities at the nodes of the curve $Y^2(z)$.
To classify the latter ones one writes
\begin{equation}
 Y^2(z)\,=\,Y_+(z)\,Y_-(z)~,
\quad\quad\quad
Y_\pm(z)\,=\,P(z)\pm Q~.
\label{pm}
\end{equation} 
Each of the two polynomials $Y_\pm(z)$ 
has $r$ nodes, call them $z_{\pm,\,i}$, $i=1,\dots r$
\begin{equation}
Y_\pm(z)\,=\,\prod_{i=1}^r\,(z-z_{\pm,\,i})\,.
\label{zpm}
\end{equation}
All the nodes $z_{\pm,\,i}$ are presumed to be enumerated in such a way as to make it certain that the pairs of the nodes $z_{\,i,\,+}$ and $z_{\,i,\,-}$ coincide in the weak coupling limit  $\Lambda \rightarrow 0$. 

\begin{figure}[tbh]
  \centering \includegraphics[ height=4.5cm, keepaspectratio=true, angle=0]{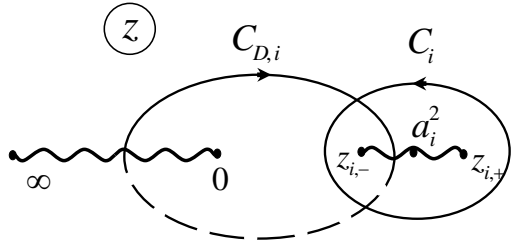}
\vspace{0cm}
\caption{ The Riemann surface for the differential $d\lambda$ in Eq.(\ref{dl}).
There are $r$ cuts between the nodes $z_{i,\,\pm}$ of the curve $Y(z)$, one 
of them is shown here; a cut between $z=0$ and $\infty$ is due to the factor $\sqrt{z}$ in $d\lambda$; solid and dashed lines show parts of $C_i$ and $C_{D,\,i}$ located on the first and second sheets of the Riemann surface, respectively. }\label{fig1} 
   \end{figure}

The analytic structure of the differential in Eq.(\ref{dl}) prompts $r$ cuts on the complex plane $z$, each cut connecting one pair of the nodes $z_{i,+}$ and $z_{i,-}$. One more cut, which runs from $0$ to $\infty$, is due to the function $\sqrt{z}$. By gluing appropriately the opposite sides of the cuts one constructs the Riemann surface for the differential, whose structure is illustrated by Fig. \ref{fig1}. For simplicity only one pair of the nodes $z_{i,\pm}$ is shown there.
   \noindent
This Riemann surface has the hyperelliptic structure and its genus $g$ equals the rank of the gauge group, $g=r$.

On this Riemann surface one defines two sets of homological cycles $C_i$ and $C_{D,\,i}$, $i=1,\dots r$, whose interception form, call it  $( \,C\,|\,C^{\,\prime}\, )$, is canonical,
see Fig. \ref{fig1},
\begin{equation}
(C_i|\,C_j)\,=\,(C_{D,i}\,|\,C_{D,j})\,=\,0~,
\quad\quad
( C_i\,|\,C_{D,j})\,=\,-( C_{D,j}\,|\,C_i )\,=\,\delta_{ij}~.
	\label{CC=d}
\end{equation}
The integrals over these cycles in Eqs.(\ref{Adl}) give the periods of the differential $d\lambda$, which are identified with the fields in (\ref{Adl}.

It was verified in \cite{Kuchiev:2009ez} that the model reproduces 
the chiral invariance and Weyl symmetry, satisfies the duality condition, and provides sensible asymptotic conditions at weak and strong coupling. 
Importantly it also reproduces a condition, whose necessity was discussed in Ref. \cite{Kuchiev:2008mv}. It was argued there that each dyon 
with electric and magnetic charges from the set $\{\CG=(\av_i, -m\av_i), ~i=1,\dots r,~m=0,1\dots \hv-1\}$, where $\av_i$ are simple coroots, can be made massless. 
Overall there are $r\hv$ dyons in this set,
and  for each of them there should exist the scalar field, which reduces the mass of this
particular dyon to zero. The compliance with this condition distinguishes this model from a number of the previously suggested  models, see discussion in \cite{Kuchiev:2008mv}.

The present work emphasizes that it is vitally important that the theory complies with $S_r$ symmetry of permutations of the sets $A_i$ and $A_{D,\,i}$, see (\ref{P}), (\ref{Pstrong}). 
It is satisfying therefore that the model fully complies with this new condition. 
To see this point observe that Eqs.(\ref{Adl})-(\ref{f2})
remain invariant under a permutation $\mathcal{P}\in S_r$, which simultaneously transforms  $A\rightarrow A^{\,\prime}$ and $A_D\rightarrow A^{\,\prime}_D$ in accord with (\ref{P}) and (\ref{Pstrong}) and in addition permutes parameters in (\ref{Adl})-(\ref{f2}) 
$a_i\rightarrow a^{\,\prime}_i$, 
where 
\begin{equation}
 a^{\,\prime}_i\,=\,\sum_{j=1}^r\,\mathcal{P}_{ij}\,a_j~.
\label{eq:}
\end{equation} 
From a technical point of view the compliance of the model with the $S_r$ symmetry 
arises due to the following three reasons. First, its variables are chosen as $A_i$ and $A_{D,\,i}$, which represent the central charges (\ref{ZiZDi}). Second, all parameters in  Eqs. (\ref{Adl}) - (\ref{f2}) can be expressed via symmetric polynomials of $a_i^2$. Third, the structure of the differential $d\lambda$ does not depend on the Cartan matrix, i. e. incorporates no details related to the geometrical structure of the lattice $\mathbb Q^\vee$.

Consider now Eqs. (\ref{AD=dF/dA}), which allow one to recover the superpotentials
$\CF(A)$ and $\CF_D(A_D)$ from the fields $A$ and $A_D$. Since we established that
the structure of the model is invariant under $S_r$ we recover from (\ref{AD=dF/dA}) that
the superpotentials $\CF(A)$ and $\CF_D(A_D)$ produced  by the model, are 
presented by functions symmetric under permutations of there arguments. This agrees with the prediction of Eq.(\ref{Fsym}). Thus the model fully reproduces the necessary general property of the theory. Namely, it is invariant under $S_r$ and it is ensured that this symmetry is implemented in such a way that the superpotentials $\CF(A)$ and $\CF_D(A_D)$ are symmetric functions of their arguments $A_i$ and $A_{D,\,i}$.

Let us reiterate the important point. The model of \cite{Kuchiev:2009ez} simultaneously complies with two conditions. One of them \cite{Kuchiev:2008mv} insists that the model should guarantee that each dyon chosen from the specific set of $r\hv$ dyons could become massless. Another, advocated in the present work, suggests that the model should implement the $S_r$ symmetry. Thus the seemingly different perspectives on the problem developed in these works agree, which is gratifying.

\section{Fields aligned along Weyl vector}
\label{Massless-monopoles}

As mentioned in the previous Section the model considered is invariant under permutations from $S_r$. Consequently, the model should reproduce the Weyl vector alignment (\ref{Weyl}).
Let us verify this fact and derive from the model the values for the coefficients $k$ and $k_D$ in (\ref{Weyl}).

Consider the scalar field aligned with the Weyl vector. Such  alignment implies that the coefficients $A_i$ in (\ref{ADi}) are all $i$-independent. Figure \ref{fig1} shows that in order to satisfy the later condition the roots of the polynomials $Y_+(z)$ and $Y_-(z)$, which are called $z_{i,\,+}$ and $z_{i,\,-}$ in (\ref{pm}), need to be $i$-independent as well
\begin{equation}
z_{i,\,+}=z_+~, \quad\quad z_{i,\,-}=z_-~, 
\label{zz}
\end{equation}
where $z_\pm$ are two $i$-independent parameters. 
From  Fig. \ref{fig1} we observe that this condition ensures also that an additional phenomenon takes place, the coefficients $A_{D,\,i}$ become also $i$-independent.
This indicates that the dual field proves to be also aligned along the Weyl vector. 
Thus the alignment of the scalar field along the Weyl vector 
compels the dual field to be aligned along this vector as well, in accord with (\ref{Weyl}).

Our next goal is to find parameters, which make Eqs.(\ref{zz}) valid.
Consider the conventional symmetric polynomials $I_k$, $k=1,\dots r$, of the variables $a_j^2$, which appear in (\ref{P(z)}), 
\begin{equation}
I_k\,=\,\sum_{j=1}^r\,a_j^{2k}~.
\label{Ik}
\end{equation}
The coefficients of the polynomial $P(z)$ in front of the term $\propto z^p$,  $1\le p\le r-1$,  can be expressed as homogeneous functions of $I_k$ with $k\le p$. For example, the coefficient in front of $z^{r-1}$ equals obviously $-(a_1^2+\dots+a_r^2)=-I_1$. Similarly, the coefficient in front of $z^{r-2}$ equals $a_1^2 a_2^2+\dots +a_{r-1}^2 a_{r}^2=(I_1^2-I_2)/2$, {\em etc}.
In order to satisfy Eq.(\ref{zz}) we  eliminate firstly the coefficients of the polynomial $P(z)$ in front of the terms $z^p$ with $1\le p\le r-1$. For this purpose it is sufficient 
to satisfy conditions
\begin{equation}
I_k\,=\,0~,\quad\quad 1\le k\le r-1~,
\label{Ik0}
\end{equation}
which can always be done by an appropriate choice of $a_j$. Namely, take $a_j$ as follows
\begin{equation}
a_j\,=\,\varkappa_j\,\lambda~,
\label{akc}
\end{equation}
where $\lambda$ is a constant, and $\varkappa_j$ are defined by the phase factors
\begin{equation}
\varkappa_j\,=\,\exp\left({i \,\pi  m_j\, }/{r}\right)~,
\label{km}
\end{equation}
in which a set $(m_1,m_2,\dots m_r)$ is an arbitrary permutation of the sequence  $(0,1,\dots r-1)$, the set with $m_j=j-1$ gives an example. It is easy to verify that thus introduced phase factors satisfy identities
\begin{equation}
\sum_{j=1}^{r}\,(\,\varkappa_j\,)^{\,2k}\,=\,r \,\delta_{\,k,\,r}\,,
\label{kp}
\end{equation}
and therefore $a_j$ from Eq.(\ref{akc}) satisfy conditions (\ref{Ik0}). Consequently, for these values of $a_j$ the polynomial $P(z)$ is reduced to 
\begin{equation}
P(z)=z^r+(-1)^r\,\prod_{j=1}^r a_j^{2}=z^r-\lambda^{2r}~.
\label{Pz-}
\end{equation}
It is taken into account here that Eqs.(\ref{akc}),(\ref{km}) imply that 
$\prod_{j=1}^r a_j^{2}=\lambda^{2r} \exp\big(\,(2\pi i/r)\sum_{j=0}^{r-1}j\,\big)=(-1)^{r-1}\lambda^{2r}$.

Equation (\ref{Pz-}) allows one to simplify the polynomials $Y_\pm(z)$ from Eq.(\ref{pm}) 
\begin{equation}
Y_\pm(z)\,=\,z^r- \lambda^{2r}\pm
[\,a \,]^{2r-\hv}\Lambda^{\hv}
\,=\,z^r -\lambda^{2r}\big( \,1 \mp ({\Lambda}/{\lambda})^{\hv}\,\big)~.
\label{Ypm}
\end{equation}
Here we resolved Eqs.(\ref{f2}),(\ref{akc}),(\ref{km}) by taking
\begin{equation}
[\,a \,]\,=\,\lambda~.
\label{lar}
\end{equation}
Substituting (\ref{Ypm}) into (\ref{Adl}) one finds that
\begin{equation}
A_i\,=\,\frac{r}{\pi }\,\int_{z_{i,\,+} }^{z_{i,\,-} }
\bigg[\,\frac{  z^{2r-1} }
{ \big( \lambda^{2r}( 1 + ({\Lambda}/{\lambda})^{\hv})-z^r \big)
\big( z^r - \lambda^{2r}( 1 - ({\Lambda}/{\lambda})^{\hv}) \big) }\, \bigg]^{1/2}
dz~.
\label{AjlL}
\end{equation}
Here the integration limits satisfy
\begin{equation}
(z_{i,\,\pm})^r\,=\,\lambda^{2r}\,\big( \,1 \mp ({\Lambda}/{\lambda})^{\hv}\,\big)~.
\label{zjpm}
\end{equation}
Resolving these conditions as follows
\begin{equation}
z_{i,\,\pm}\,=\,
\lambda^{2}\,\big( \,1 \mp ({\Lambda}/{\lambda})^{\hv}\big)^{1/r}~,
\label{zjpmlambda}
\end{equation}
we fulfil the goal set in (\ref{zz}), the found $z_{i,\,\pm}$ are $i$-independent.
To simplify notation it is convenient to introduce a constant $\nu$ and parameter $u$ 
\begin{align}
&\nu\,=\,1/(2\,r)~,
\label{nu}
\\
&u\,=\,\left(\,{\lambda}/{\Lambda}\,\right)^{\hv}~,
\label{u}
\end{align}
and  change the integration variable in Eq.(\ref{AjlL}) from $z$ to $x$ as follows
\begin{equation}
z\,=\,\lambda^2\,\big(1-{x}/{u}\,\big)^{2\nu}~.
\label{zx}
\end{equation}
After that Eq.(\ref{AjlL}) is reduced to
\begin{align}
&A_i\,=\,K_\nu(u) \, \lambda\,,
\label{ab}
\\
&K_\nu(u)\,=\,\frac{1}{\pi }\,\int_{-1}^{1} 
\left( 1-\frac{x}{u} \right)^{\nu} \frac{dx}{\sqrt{1-x^2}   }\,=\,
\left(1+\frac{1}{u}\right)^\nu\,F\left(\,\frac{1}{2},\,-\nu,\,1,\,\frac{2}{1+u}\,\right)\,.
\label{Ku}
\end{align}
Here $F(a,b,c,z)$  is the hypergeometric function. 
Similar calculations allow one to find the dual field 
\begin{align}
&A_{D,\,i}\,=\,K_{D,\,\nu}(u) \,\lambda~,
\label{aDb}
\\
&K_{D,\,\nu}(u)\,=\,\frac{i}{\pi }\,\int_{1}^{u} 
\left( 1-\frac{x}{u} \right)^{\nu} \frac{dx}{\sqrt{x^2-1}   }
\label{KDu}
\\
&\quad\quad\quad=
\frac{i} {\sqrt{2}~\pi}~
B(\,1+\nu,\,{1}/{2}\,)
\,{u^{-\nu}}\,(u-1)^{1/2+\nu}
\,{F}\left(\,\frac{1}{2},\,\frac{1}{2},~\frac{3}{2}+\nu,\,
\frac{1-u}{2}\,\right).
\nonumber
\end{align}
Here $B(x,y)$ is the Euler beta-function.
An alternative representation for $K_\nu(u)$ and $K_{D,\,\nu}(u)$ is discussed in Appendix \ref{Alternative}.

Substituting $A_i$, $A_{D,\,i}$ from Eqs.(\ref{ab}),(\ref{aDb}) to Eqs.
(\ref{ADi}) we can conveniently present the found fields $A$ and $ A_\text{\it D}$ as follows
\begin{align}
&A~~=~K_\nu(u)\,\lambda\,\rho~,
\label{A=Ka}
\\
& A_\text{\it D}\,=\,K_{D,\,\nu}(u)\,\lambda\,\rho~,
\label{AD=KDa}
\end{align}
where, remember, $u$ is a function of $\lambda$ defined in (\ref{u}). Consequently the found $A$ and $ A_\text{\it D}$ are functions of this parameter as well.

Equations (\ref{A=Ka}), (\ref{AD=KDa}) state that $A$ and $ A_\text{\it D}$ are aligned along the Weyl vector $\rho$, i. e. reproduce the Weyl vector alignment predicted in (\ref{Weyl}). An advantage of the present consideration is that it provides explicit expressions for the coefficients $k=K_\nu(u)\lambda/\Lambda$ and $k_\text{\it D} =K_{D,\,\nu}(u)\lambda/\Lambda$, which specify 
the aligned fields $A$ and $ A_\text{\it D}$ completely.

It is easy to verify that Eqs.(\ref{A=Ka}),(\ref{AD=KDa}) comply with the Seiberg-Witten solution for the gauge theory with the simplest SU(2) gauge group \cite{Seiberg:1994rs}. Taking for this group the values $r=1$, $\hv=2$, $\rho={1}/{\sqrt 2}$,
one finds using Eqs.(\ref{Ku}) and (\ref{KDu}) 
\begin{align}
A=\frac{\Lambda}{\pi\sqrt 2}\,\int_{-1}^1\left(\frac{u-x}{1-x^2}\right)^{\!1/2}\!dx~,
\quad\quad
 A_\text{\it D}=i\,\frac{\Lambda}{\pi\sqrt 2}\,\int_{1}^u
\left(\frac{u-x}{x^2-1}\right)^{\!1/2}\!dx~,
\label{KSW}
\end{align}
which reproduce the Seiberg-Witten solution.

Returning to the case of an arbitrary gauge group let us verify the weak coupling limit. From 
Eqs.(\ref{ab}) - (\ref{KDu}) one obtains for  $|\lambda|\gg \Lambda$
\begin{align}
&A\,~~\approx~\lambda\,\rho~,
\label{Aa}
\\
& A_\text{\it D}\,\approx~\frac{i }{2\pi}\,\hv \lambda\,\rho\,
\Big(\,\ln\frac{\lambda^2}{\Lambda^2}+
\frac{2}{\hv}\,\big(\,\ln 2-\psi (1+\nu)-C\,\big)\, \Big)~.
\label{ADa}
\end{align}
Here $C\simeq 0.577$ is the Euler constant and $\psi(x)=\Gamma^{\,\prime}(x)/\Gamma(x)$ is the digamma function. We see that Eqs.(\ref{Aa}),(\ref{ADa}) comply with the asymptotic behavior of the fields $A$ and  $ A_\text{\it D}$ at weak coupling (\ref{ADPT}). 
As an illustration take the SU(2) gauge group  where Eq.(\ref{ADa}) yields
$ A_\text{\it D}\,\approx\, i\, (A/\pi)\,(\,\ln\,(A^2/\Lambda^2)+4\ln2-2\,)$,
which agrees with the asymptotic relation that follows from Eqs.(\ref{KSW}) for $|u|\gg 1$.

Deriving Eq.(\ref{ADa}) it is convenient to rely on the known identity $F(a,b,c,w)=(1-w)^{-\alpha}\,{F}(a,c-b,c,w/(w-1))$
expressing the hypergeometric function from Eq.(\ref{KDu}) via the hypergeometric function of the argument $z=(u-1)/(u+1)$, and then apply to the later function the asymptotic relation
\begin{equation}
F(a,\,b,\,a+b,\,z)\,\approx\,-\frac{1}{B(a,b)}\,\big(\,\ln(1-z)+\psi(a)+\psi(b)+2C\,\big)~,
\label{latter}
\end{equation}
which is valid when $z\approx 1$, as it follows from Eq.(15.3.10) of Ref. \cite{Abramowitz:1972}.
Asymptotic properties of Eqs.(\ref{ab}), (\ref{aDb}) in the opposite limit of strong coupling are discussed in Appendix \ref{asymptotic}.


One more useful verification of Eqs.(\ref{A=Ka}), (\ref{AD=KDa}), which is valid at arbitrary coupling, stems from the discrete chiral transformations. Let us make in the integrals
in Eqs. (\ref{Ku}) and (\ref{KDu}), which define the coefficients
$K_{D}(u)$ and $K_{D,\,\nu}(u)$, the transformation $\lambda\rightarrow \exp(2i\gamma)\lambda$, where the phase $\gamma$ runs continuously from  $\gamma=0$ to $\gamma=\pi/(2\hv)$. Our aim is to emulate the chiral transformation described by  Eqs. (\ref{AgA}),(\ref{ADct}) for $m=1$. The transformation of $\lambda$ implies the corresponding variation for $u$, $u\rightarrow \exp(\,2\, i\gamma \hv)\,u$, which changes this parameter from its initial value $u$ to $-u$. From (\ref{Ku}) and (\ref{KDu}) one observes that under this transformation $K_\nu(u)$ does not change, $K_\nu(-u)=K_\nu(u)$, while $K_{D,\,\nu}(u)$ exhibits the following variation $K_{D,\,\nu}(-u)=K_{D,\,\nu}(u)-K_\nu(u)$. Combining the later fact with the corresponding variation of $\lambda$, $\lambda\rightarrow \exp(\pi i/\hv)\lambda $, we find that the 
the chiral transformation of fields $A$ and $ A_\text{\it D}$ in Eqs. (\ref{A=Ka}) and (\ref{AD=KDa}) comply with Eq.(\ref{ADgamma}), as necessary.
The possible issues related to the absence  of the Weyl vector alignment in some previously suggested  models implementing the Seiberg-Witten approach are discussed in Appendix \ref{Comparisons}.

Summarizing, Eqs.(\ref{A=Ka}),(\ref{AD=KDa}) give explicit description of the fields 
$A$ and $ A_\text{\it D}$ when they are aligned along the Weyl vector. 


\section{Massless monopoles and dyons}
\label{Massless-dyons}

Consider the state of the supersymmetric $\CN=2$ gauge theory, in which the dual field turns zero  and correspondingly all monopoles become massless
\begin{equation}
 A_\text{\it D}\,=\,m_\text{\,mon}\,=\,0~.
\label{zer}
\end{equation}
Equations (\ref{KDu}) and (\ref{AD=KDa}) show that this condition takes place provided 
\begin{equation}
u\,=\,1~.
\label{u=1}
\end{equation}
Correspondingly, resolving Eqs. (\ref{u}), (\ref{u=1})  
for the parameter $\lambda$ we find
\begin{equation}
\lambda\,=\,\Lambda~.
\label{lL}
\end{equation}
There is no need to consider here a more general solution, which includes a phase factor $\lambda=\exp(2\pi n \,i/\hv)\Lambda$, $n\in \mathbb Z$, since the cutoff parameter $\Lambda$ originally appears in the theory as the power $\Lambda^{\hv}$,  see (\ref{P(z)}),  and hence can always be redefined with the help of the same phase factor, $\Lambda \rightarrow \Lambda^{\,\prime}=\exp(2\pi n \,i/\hv)\Lambda$.

From Eqs. (\ref{lL}) and (\ref{A=Ka}), (\ref{Ku}) we find the scalar field $A$, which 
makes (\ref{zer}) valid
\begin{align}
&A\,=\,k_\nu\,\Lambda\,\rho~.
\label{Af}
\\
&k_\nu\,=\,K_\nu(1)\,=\,\frac{2^{\,\nu}}{\pi }\,\int_{0}^1x^{\nu-1/2}\frac{dx }{\sqrt{1-x}}\,=\,
\,\frac{2^{\,\nu}}{\pi }\,B(\,1/2+\nu,\,1/2\,)~.
\label{k}
\end{align}
Equation (\ref{Af}) states that when the dual field is absent, 
$A_\text{\it D}=0$, the scalar field is necessarily aligned along the Weyl vector. This is in line with the general, model independent statement made in (\ref{corollary}).
Eq.(\ref{k}) goes further specifying a precise value of the numerical coefficient $c=k_\nu$
in this equation. The asymptotic behavior of the fields aligned with the Weyl vector in the vicinity of the critical point $u=1$  is discussed in Appendix \ref{asymptotic}.

Similarly one considers a situation when all the dyons with charges $\CG_\text{dyon}=(g,-mg)$, where $m\in \mathbb Z$ is fixed and $g$ is an arbitrary positive vector in the lattice of coroots, are massless.  Remembering discussion related to Eqs.(\ref{Pho0}) - (\ref{Phim=}) one concludes that this situation is covered by Eqs.(\ref{Phim=}) in which 
\begin{equation}
k_\text{mon}\,=\,k_\nu~.
\label{kmon-knu}
\end{equation}

\section{Summary and discussion}
\label{Summary and discussion}

The Dirac-Schwinger-Zwanziger quantization ensures that the electric and magnetic charges 
available in the theory are pinned to the vertexes of the lattice $\mathbb Q^\vee$, see (\ref{g=ng}), located in the Cartan algebra of the gauge group.
It is demonstrated that the presence of this lattice makes the theory invariant under the discrete group of permutations $S_r$, where $r$ is the rank of the gauge group. 
To interpret this symmetry condition in simple physical terms one can say that the theory is invariant under a simultaneous permutation of $r$ simple electric charges and $r$ simple monopoles ($r$ simple electric charges and $r$ simple monopoles have the electric or magnetic charges, which equal $r$ simple coroots of the Cartan algebra).


The symmetry of the theory under $S_r$ makes the superpotential $\CF(A)$ 
a symmetric function of $r$ central charges, which correspond to the simple electric charges,
see (\ref{Fsym}). As a result the $\tau$-matrix of coupling constants is transformed under $S_r$ in a simple, appealing way (\ref{Ptau}). These properties manifest themselves transparently when the scalar field $A$ and its dual $A_D$ are written in the basis of fundamental weights of the Cartan algebra (\ref{ADi}). In this approach the two sets of their expansion coefficients $A_i$ and $A_{D,\,i}$ equal the central charges for simple electric charges and simple monopoles respectively (\ref{ZiZDi}). These two sets of central charges obey Eqs.(\ref{AD=dF/dA}), which relate them to the superpotential.

An interesting implication of the symmetry under $S_r$ is the Weyl vector alignment (\ref{Weyl}). When expressed in simple geometrical terms this condition
states that if one of the two vectors, either $A$ or $A_D$ is chosen to be collinear to the Weyl vector $\rho$ of the Cartan algebra, then the other vector is necessarily aligned along the same vector as well, which makes all three of them collinear $A\parallel \rho \parallel A_D$.
In physical terms this condition means that if the system is driven into a state, in which  central charges for $r$ simple electric charges are equal, then central charges of $r$ simple monopoles necessarily turn equal as well, and vise versa, equality between central charges of simple monopoles makes central charges of simple electric charges equal. The identity between central charges implies degeneracy of masses. Thus, when the Weyl vector alignment $A\parallel \rho \parallel A_D$ takes place, then the masses of $r$ simple electric charges are degenerate, and at the same time the masses of $r$ simple monopoles are degenerate as well. These properties fit very well within the general physical picture associated with the idea of duality and can be considered a good test for the theory.

A simple and very important corollary arises for the state where all monopoles are massless. In this case the Weyl vector alignment ensures that the scalar field is necessarily aligned along the Weyl vector and therefore its value is completely specified by (\ref{corollary}).
In physical terms this means that if all monopoles are massless, then the central charges of all simple electric charges are equal. Consequently the masses of simple electric charges 
are equal as well. It is interesting that the electric charges exhibit such a special behavior at this important point. 

It is shown in Section \ref{Comparisons} that the model suggested previously for SU($n$) gauge groups (and probably a number of other models developed for other gauge groups prior to \cite{Kuchiev:2009ez}) does not comply with the Weyl vector alignment and therefore is certainly different from the model of \cite{Kuchiev:2009ez}.
This work suggests a general, model independent and calculation independent way
by which this issue can be addressed and resolved. It is argued that since there exists a discrete symmetry in the theory the discrepancy between different approaches can be traced down to different discrete symmetry groups, or different representations of these groups, on which each theoretical framework is developed. It is vital that the symmetry group keeps the set of central charges $\mathfrak Z$ invariant.
The present work shows that this condition makes the superpotential invariant under $S_r$. 
It would be interesting to study the discrete groups and their representations in the previously developed models. 

The results of the present work are compared with the model of \cite{Kuchiev:2009ez}, which extended the Seiberg-Witten approach to the supersymmetric $\CN=2$ theory for an arbitrary gauge group. It is shown that this model complies with $S_r$ symmetry. The model is used to find explicitly the behavior of the scalar and dual fields when they both are aligned along the Weyl vector, see (\ref{ab}) - (\ref{AD=KDa}). In particular, the point where all monopoles are massless is identified in (\ref{Af}). The structure of the model was initially inspired by \cite{Kuchiev:2008mv}, which argued that for any dyon with charges $\CG=(\av_i,-m\av_i)$ there should exist a state of the system, in which this dyon is massless.  Thus, different starting points employed in the present paper and in  \cite{Kuchiev:2008mv,Kuchiev:2009ez} lead to the mutually consistent outcome.

The found symmetry under permutations within the specified set of dyons greatly simplifies the low-energy description of the supersymmetric $\CN=2$ gauge theory and helps to see its structure more clearly.

\vspace{0.3cm}
The financial support of the Australian Research Council is acknowledged.

\appendix

\section{Comment on weak coupling}
\label{subtlety}

Continuing discussion in Section \ref{Weak coupling} note that 
one can recover the important Eq.(\ref{ADPT}) from the superpotential
using two quite different techniques.
Following the approach of the present work one takes the superpotential
$\CF(A)$ from (\ref{Fweak}), differentiates it over the variable $A_i$, and using (\ref{AD=dF/dA}) reproduces (\ref{ADPT}). 
In a more traditional formalism one would take $\CF_0(A)$ from (\ref{Ftrad}), differentiate it over $A^\perp_s$ presuming that the result reproduces
the dual variable in the orthogonal basis, $A^\perp_{D,s}=\partial \CF(A)/\partial A^\perp_s$. This chain
of transformations goes against the main argument of the present work. However, if one allows oneself to accept these transformations, then  the correct Eq.(\ref{ADPT}) is reproduced. Trouble is that the success in derivation of (\ref{ADPT}) from $\CF_0(A)$ is accidental. For example, cultivating this disputed in the present work line of arguments one would be led to believe that the matrix of coupling constants is diagonal in the orthogonal basis $\tau_{0,\,st}=\partial A^\perp_{D,s}/\partial A^\perp_t\approx \delta_{st}(i \hv/2\pi )\ln(A^2/\Lambda^2)$. This outcome contradicts Eq.(\ref{tauPThOrt}) and hence in accord with the arguments of the present paper is misleading.

\section{Alternative form for hypergeometric functions}
\label{Alternative}

Quadratic transformation formulas for the hypergeometric functions, see Section 15.3 of \cite{Abramowitz:1972},  give another option for presenting integrals in Eqs.(\ref{Ku}), (\ref{KDu})
\begin{align} 
&K(u)\,=\,
F\left(\,\frac{1}{2}-\frac{\nu}{2},\,-\frac{\nu}{2},\,1,\,\frac{1}{u^2}\,\right)~,
\label{KuA}
\\
&K_D(u)=
\frac{i\,B(1+\nu,\,1/2)}{2^{1+\nu}\,\pi}\,
\left(1-\frac{1}{u^2}\right)^{1/2+\nu}
\!{F}\left(\frac{1}{2}+\frac{\nu}{2},\,1+\frac{\nu}{2},\,\frac{3}{2}+\nu,\,1-\frac{1}{u^2}\right).
\label{KDuA}
\end{align}
Being well defined at $u>1$ they require accurate handling outside this region.

\section{Asymptotic properties at strong coupling}
\label{asymptotic}
Consider the limit $\lambda\rightarrow \Lambda$, in which the monopoles become massless and the strong coupling approximation is valid. It is  convenient in this case to distinguish the theory based on the SU(2) gauge group from theories with more complex gauge groups. The SU(2) group has rank $r=1$, and correspondingly there is only one monopole in the theory. Any other gauge group has a larger rank $r>1$, and hence there are  several monopoles there. When the scalar field $A$ is tuned is such a way that only one monopole is light, then $A$ possesses the well known strong-coupling logarithmic singularity. To illustrate this fact remember what this singularity looks like for the theory based on the SU(2) gauge group \cite{Seiberg:1994rs} when Eqs. (\ref{KSW})  give
\begin{equation}
A\,\approx\,\Big(\,\frac{2}{\pi}-\frac{1}{4\pi}(u-1)\ln(u-1)\,\Big)\,\Lambda~,
\quad\quad\quad
A_\text{\it D}\,\approx\,\frac{i}{4}\,(u-1)\,\Lambda~,
\label{Au-1}
\end{equation}
where $u=\lambda^2/\Lambda^2=2a^2/\Lambda^2\approx  1$. Similarly, the logarithmic singularity manifests itself for the theory based on any other gauge group. However, such  singularity is present provided only one monopole is massless. When several monopoles become massless simultaneously, which is possible in the theories based on gauge groups with $r>1$,  the situation is different. In that case several logarithmic singularities related to different  monopoles can collide changing the nature of the singularity. 
Eqs.(\ref{A=Ka}),(\ref{AD=KDa}) illustrate the later statement for the particular situation, when all the fields are scaled along the $\rho$ direction in the Cartan subalgebra.
Assume that $r>1$ and consider the region $u\approx 1$. Expressing the hypergeometric function of the argument $z$ via two hypergeometric functions with the argument $1-z$ one finds that in this case
\begin{align}
&A~~\,\approx\,\big(\,k_\nu+k^{\,\prime}_\nu\,(u-1)^{1/2+\nu}\,\big)\,\lambda\,\rho~,
\label{Asc}
\\
& A_\text{\it D}\,\approx\,i~k_\nu^{\,\prime}\,\cot\pi\nu\,\,(u-1)^{1/2+\nu}\,\lambda\,\rho~,
\label{ADsc}
\end{align}
where the constant $k_\nu$ is defined in Eq.(\ref{k}), while $k_\nu^{\,\prime}$ is a new constant
\begin{equation}
k_\nu^{\,\prime}\,=\,
\frac{1}{\sqrt{2}~\pi }~B(-1/2-\nu,1/2)~.
\label{k'}
\end{equation}
Deriving (\ref{ADsc}) an identity $B(1+\nu,1/2)=\cot\pi\nu \,B(-1/2-\nu,1/2)$ was employed.

\section{Comparison with other models}
\label{Comparisons}

Let us compare the results of Section \ref{Degeneracy-Weyl} with the previously known low-energy solutions of the $\CN=2$ supersymmetric gauge theory. Take first the solution for the unitary SU($r$+1) gauge theory suggested in 
\cite{
Klemm:1994qs,
Argyres:1994xh
}.
The scalar field in this solution has the following form
\begin{equation}
A^\text{\,SU(r+1)}\,=\,\sum_{k=1}^{r+1} A^{\,\perp}_k\,\varepsilon_k~.
\label{perp}
\end{equation}
Here $r$+1 vectors $\varepsilon_k$ constitute the orthonormal basis in the ($r$+1)-dimensional space, $\varepsilon_k\cdot \varepsilon_l=\delta_{kl}$, while the coefficients $A^{\,\perp}_k$ satisfy condition $\sum_{k=1}^{r+1}A^{\,\perp}_k=0$, which leaves only $r$ of them independent. 
(To avoid confusion note that the basis in Eq.(\ref{perp}) is built from $r+1$ vectors $\varepsilon_k$ in $r+1$ dimensional space while previously in (\ref{perp0}) the orthogonal basis included $r$ vectors $\varepsilon_s$ in $r$-dimensional space.) 

It was stated in \cite{Douglas:1995nw} (see Eq.(2.11) of \cite{Douglas:1995nw}, which is written in terms of $a_k=\sum_{i=1}^k A^{\,\perp}_i$) that at the point 
where the dual field vanishes, $ A_\text{\it D}=0$,  the expansion coefficients of the scalar field in (\ref{perp}) satisfy
\begin{equation}
A^{\,\perp}_k\,=\,\frac{2\sqrt{2}\,\Lambda\,(r+1)}{\pi}\,\Big(\,\sin\frac{\pi k}{r+1}-\sin\frac{\pi (k-1)}{r+1}\, \Big)\,\propto\,
\cos\frac{\pi\,(k-1/2)}{r+1}~.
\label{SUN12}
\end{equation}
Remember that the Weyl vector for the SU$(r+1)$ group equals 
\begin{equation}
2\rho\,=\,r\varepsilon_1+(r-2)\varepsilon_2+\,\dots\,-(r-2)\varepsilon_r-r\varepsilon_{r+1}~.
\label{2rho}
\end{equation}
Observe that for $r\ge 3$ the scalar field defined in Eq.(\ref{SUN12}) is not collinear with the Weyl vector (\ref{2rho}) because, 
for example, $A_2^\perp/A_1^\perp\ne (r-2)/r$. Thus the scalar field does not comply with
the corollary to the Weyl vector alignment Eq.(\ref{corollary}). Hence  it contradicts the Weyl vector alignment (\ref{Weyl}) itself. 
Preliminary examination of other suggested previously solutions for SO$(n)$ and  Sp$(2r)$ gauge theories gives a similar outcome, they seem to not comply with the Weyl vector alignment, which presents an important issue.
Another, though related issue stems from the fact that it is generally believed that the available instanton calculations comply with the mentioned model for the SU($n$) gauge group (and also reproduce the results of other previously known models). As a result 
these calculations should contradict the Weyl vector alignment as well.

This work suggests a model independent and calculation independent way by which this issue can be addressed. It is argued that there exists a discrete symmetry group $S_r$ in the theory, 
which keeps the superpotential invariant, see (\ref{Fsym}).
As a result it also keeps invariant the set of central charges $\mathfrak Z$.
From a different perspective, the discrete symmetry is usually introduced into the theoretical description at the moment when some particular model is formulated. For each model it is possible therefore to specify the discrete symmetry group, compare it with $S_r$, and also verify if it is represented properly, i. e. keeps invariant the set of central charges  $\mathfrak Z$. The fact that the Weyl vector alignment is violated may probably be considered an indication that either the discrete group is not $S_r$, or an issue exists with the way $S_r$ is represented in the given model. (Similarly, it would be interesting to realize which discrete symmetries and which representations are incorporated in the corresponding instanton calculations.)


\end{document}